\documentclass{elsart}

\usepackage{amsmath,amssymb,bbm}
\usepackage{epsfig}

\begin{document}

\begin{frontmatter}
\begin{flushright}
{\bf SPHT-T04/124}\\
{\bf CU-TP-1117}\\
{\bf CPHT-RR-045.0804}
\end{flushright}
\title{Universal behavior of
QCD amplitudes at high energy
from general tools of statistical physics
}

\author[1]{E. Iancu\thanksref{th2},}
\author[2]{ A. H. Mueller\thanksref{th1},}
\author[3,4]{ S.
Munier\thanksref{th2}}

\address[1]{Service de Physique Th{\'e}orique,  Unit\'e de recherche
associ\'ee au CNRS (URA D2306),
Commissariat \`a l'\'energie atomique, 91191~Gif-sur-Yvette, France}
\address[2]{Department of Physics, Columbia University,
New York, NY 10027, USA}
\address[3]{Centre de Physique Th{\'e}orique, Unit\'e mixte de
recherche du CNRS (UMR 7644),
{\'E}cole Polytechnique, 91128~Palaiseau, France}
\address[4]{Dipartimento di Fisica, Universit\`a di Firenze, via Sansone 1,
50019~Sesto F., Florence, Italy.}

\thanks[th2]{Membre du Centre National de la Recherche Scientifique
(CNRS), France.}
\thanks[th1]{This work is supported in part by the Department of Energy.}

\begin{abstract}
We show that high energy scattering is
a statistical process essentially similar to reaction-diffusion
in a system made of a finite number of particles.
The Balitsky-JIMWLK equations correspond to the
time evolution law for the particle density.
The squared strong coupling constant plays
the role of the minimum particle density. Discreteness is related
to the finite number of partons one may observe in a given event
and has a sizeable effect on physical observables.
Using general tools developed recently in statistical physics, we
derive the universal terms in the rapidity dependence of the
saturation scale and the scaling form of the amplitude, which
come as the leading terms in a large rapidity and small coupling
expansion.
\end{abstract}

\end{frontmatter}

\section{Introduction}

Much progress has been made recently in understanding high energy
hard scattering in QCD at or near the unitarity limit.  General
equations have been given by Balitsky \cite{bal} and by
Jalilian-Marian, Iancu, McLerran, Leonidov, Kovner and Weigert
(JIMWLK) \cite{Jal,Ian,Wei} which generalize the
Balitsky-Fadin-Kuraev-Lipatov (BFKL) evolution \cite{BFKL} to
the region where unitarity (saturation) effects become important.
The Balitsky-JIMWLK equations are nonlinear operator equations,
while a ``mean field'' version of the equations, the
Balitsky-Kovchegov (BK) \cite{bal,Kov} equation, is a nonlinear
equation for the scattering amplitude and has been widely studied
recently. The scattering amplitude which emerges from the BK
equation is, in general terms, characterized by the energy
(rapidity) dependence of the saturation momentum $Q_s(Y)$
\cite{Gri,Bier,Ita,Mue,Tri,Mun}, and by geometric scaling
\cite{Kwi,Ita,Mue}, the statement that the scattering amplitude
$A(Q^2,Y)$ is equal to a function of a single variable
$A(Q^2/Q_s^2(Y))$.

However, it has not been clear to what level the general
properties of solutions to the BK equation are shared by solutions
to the Balitsky-JIMWLK equations. This is the problem we address
in this note.  Our object is to describe the energy-dependence of
the saturation momentum and the scaling properties of the
scattering amplitude which should emerge from the Balitsky-JIMWLK
equations.  When viewed in a particular way the problem here looks
identical to a class of problems studied recently in statistical
physics \cite{Bru,Van}. In the statistical physics problems  the
Fisher-Kolmogorov-Petrovsky-Piscounov (FKPP) equation \cite{KPP}
approximately describes the time evolution of certain quantities
in some discrete systems. In the QCD problem of dipole-dipole
scattering the BK equation describes the rapidity and $Q^2$
dependence of the scattering amplitude.  As has been noticed
recently, the BK equation is in the same universality class as the
FKPP equation and this fact gives a powerful and general
derivation of the energy dependence of $Q_s(Y)$ and of geometric
scaling \cite{Mun}.

The FKPP equation has limitations in applications to average
quantities in discrete  statistical systems.  When the
discreteness is not important, that is in a region where many
``objects'' are present, the FKPP equation is a good approximation
to the actual evolution of the system. However, when only a few
objects are involved discreteness effects are significant and the
FKPP description breaks down \cite{Bru,Pan}.  A similar effect
occurs in QCD evolution,  and this is seen most easily by viewing
the scattering of an elementary dipole of size  $ r$ on an evolved
dipole of initial size $r_0$ in terms of the $Y-$evolution of
particular configurations of dipoles starting from $r_0.$ (The
scattering amplitude is then given by an average over all possible
configurations.) The dipoles making up a configuration are the
discrete elements of our system. The importance of fluctuations
due to discretness in QCD evolution was first noticed by Salam
from Monte-Carlo studies \cite{Sal}. { More recently, the
importance of fluctuations has been reiterated in the context of
non-linear evolution in Ref. \cite{IM}, where the role of rare
fluctuations in the approach of the S-matrix towards the unitarity
limit has been discussed, and also in Ref. \cite{Wei,BIW}, where
JIMWLK evolution has been reformulated as a random walk in some
functional space, thus emphasizing its stochastic nature. This
last formulation lies also at the basis of the numerical studies
of JIMWLK evolution in Ref. \cite{Rum}.}

So long as the
dipole occupancy in a configuration is large compared to one, use of
the BK equation should be a
good approximation for the evolution of our configuration.  However,
when there are only a few
dipoles of size  $r$  in our particular configuration  BK evolution
cannot be expected to be accurate.
Indeed in a discrete picture occupancy can go below one only by
becoming zero which stops the evolution
along that path.  Thus, using the BK equation with a cutoff when
dipole occupancy is near one, or
equivalently when the scattering amplitude for a particular
configuration becomes of size $\alpha_s^2$,
exactly the same procedure used for discrete statistical systems
\cite{Bru,Pan},
should be a good representation of
the evolution of the system.  This cutoff is essentially the same as
that introduced in Ref.~\cite{Sho}, and
the present discussion can be viewed as a justification of the
procedure used there at least for the
calculation of the energy dependence of $Q_s(Y)$.

To also compute the dependence of the scattering amplitude upon
the dipole size, and thus compare with the geometrical scaling
form of the solution to BK equation, one needs to understand the
fluctuations  of the saturation momentum from one configuration to
another.  Here our control is less complete,
and we rely on a scaling law recently seen in numerical simulations.
The scale which
emerges is equal to the square root of the value found in
Ref.~\cite{Sho} where fluctuations at the boundary were not
included.

Finally, it should be emphasized that our description is for a
scattering at a definite impact parameter.
A more complete discussion, including impact parameter dependences,
will be given later \cite{lle}.


\section{High energy scattering as a statistical process}

We consider the scattering of a dipole of variable size $r$ (the
probe) off a dipole of size $r_0$ (the target). A natural variable
that will be used throughout is $\rho\!=\!\ln(r_0^2/r^2)$. We go
to the rest frame of the probe so that the target carries all the
available rapidity $Y$. The impact parameter $b$ between the
dipoles is fixed.

The target interacts through its quantum fluctuations, which at high
energy are dominated by gluons.  It proves useful to represent this
set of partons by color dipoles \cite{CD}.  This is possible in the
dilute regime in which saturation effects (i.e. interactions among
gluons inside the target wavefunction) are negligible, but the effects
of fluctuations should on the contrary be important. Then, the
dipole picture emerges in the large--$N_c$ limit, in which gluons are
similar to zero--size $q\bar q$ pairs and non-planar diagrams are
suppressed.  The dipole approximation breaks down when the amplitudes
approach their unitarity limits (indeed, in the considered frame, the
unitarity corrections are tantamount to saturation effects in the
target).  However, that does not hamper getting the right asymptotics
for physical quantities like the saturation scale since, as we
shall see, this is controlled by the dynamics in the tail of the
distribution at high transverse momenta, or small dipole sizes.

We denote by $T(r,r_0)$ the scattering amplitude of the probe off a
given partonic realization $|\omega\rangle$ of the target (the
dependence on $b$ is understood).  It is a random variable, whose
probability distribution is related to the distribution of the
different Fock state realizations of the target. The values of $T(r)$
range between 0 (weak interaction) and 1 (unitarity limit).  $T(r)$
will be an essential intermediate quantity in our calculations, but it
is not an observable.  The physical dipole-dipole scattering amplitude
$A(r,Y)$ is the statistical average over all partonic fluctuations of
the target at rapidity $Y$, i.e.  $A(r,Y)=\langle T(r) \rangle_Y$.

When $T$ is small, $T(r,r_0)=\sum T_{el}(r,r_i)$, where $i$ labels the dipoles
in the Fock state of the target at the time of the interaction.
$T_{el}$ is the elementary dipole interaction and
is essentially local in impact parameter.
$T_{el}$ behaves like
\begin{equation}
T_{el}(r,r_i)\sim
\alpha_s^2 \,\frac{r_<^2}{r_>
^2}\ ,\ \mbox{with}\
r_<=\min(|r|,|r_i|)\ ,\ r_>=\max(|r|,|r_i|)
\label{Tel}
\end{equation}
when the dipoles overlap, and vanishes otherwise (see the insert in
Fig.~1). We have neglected ${\mathcal{O}}(1)$ factors and
logarithms, but these approximations do not affect the results that
we shall obtain, which are largely independent of the details.
Eq.~(\ref{Tel}) shows that the amplitude $T(r,r_0)$ is simply
counting the number $n(r,r_0)$ of dipoles of size $r$ within a disk
of radius $r$ centered at the impact parameter of the external
dipole (the dipole occupation number):
\begin{equation}\label{ndef}
T(r,r_0)\sim \alpha_s^2\, n(r,r_0)\ .
\end{equation}
Note that $n(r,r_0)$ can take only discrete values. Thus, in this
description, fluctuations in $T$ emerge naturally as fluctuations in
the {\it particle} (here, dipole) {\it number}, which should be
especially important in the regime where $n\sim{\mathcal O}(1)$.
The unitarity bound on $T$ implies that $n(r,r_0)$ is also constrained
by an upper bound $N\sim{\mathcal O}(1/\alpha_s^2)$.

The dipole picture also provides us with the evolution law for the
dipole distribution with increasing rapidity. Consider a small increment
$dY$ of the total rapidity from a boost of the target.
Then, each of the dipoles $r$ already present in the wave
function from the previous evolution and for which $n(r,r_0)\ll1/\alpha_s^2$
may split into two new dipoles, of respectives sizes $z$ and $r-z$,
with a differential probability\cite{CD}
\begin{equation}
dP=dY\frac{\bar\alpha}{2\pi}\frac{r^2}{z^2(r-z)^2}d^2z
=[\lambda\,\bar\alpha dY] \times [p(z,r\!-\!z|r)\,d^2z]
\label{splitting}
\end{equation}
where $\bar\alpha=\alpha_s N_c/\pi$. We see that
$\bar\alpha Y$ is the natural evolution variable: we will call it ``time''.
The second equality in Eq.~(\ref{splitting}) expresses $dP$
as the product of the inclusive probability of splitting\footnote{%
An ultraviolet cutoff is understood in Eq.~(\ref{lambda}). It disappears in
physical quantities.}
\begin{equation}
\lambda=
\int \frac{d^2z}{2\pi} \frac{r^2}{z^2(r\!-\!z)^2}
\label{lambda}
\end{equation}
in the time interval $\bar\alpha\,dY$, by the
conditional distribution of the sizes of the produced dipoles
\begin{equation}
p(z,r-z|r)\,d^2z=
{\lambda}^{-1}\frac{r^2}{z^2(r\!-\!z)^2}\,\frac{d^2z}{2\pi}\ .
\label{p}
\end{equation}
By applying Eqs.~(\ref{splitting})--(\ref{p}) to any of the dipoles
$r_i$ present in the wavefunction, one can easily deduce the evolution
law for the dipole configuration as a whole. This leads to a
description of the (dilute tail of the) target wavefunction as a
stochastic ensemble of dipole configurations endowed with a
probability distribution which evolves with $Y$ according to a master
equation \cite{IM04}. This picture, which is similar to certain
problems in statistical physics, is particularly appropriate for a
study of fluctuations, since the discreteness of the dipole number
is explicit.
However, this picture breaks down, as anticipated, when the dipole
occupation numbers --- which in the dilute regime rise exponentially
with $Y$ --- become of ${\mathcal O}(1/\alpha_s^2)$, and saturation
effects start to play a role. But in this high density regime, one can
rely on a different formalism, the color glass condensate \cite{Ian},
which is an effective theory for gluon correlations in the target
wavefunction at small $x$, and is endowed with a {\it functional} evolution
equation --- the JIMWLK equation \cite{Jal,Ian,Wei} --- which shows how 
these correlations change under a boost. When this formalism is applied 
to the scattering between the color glass (the target) and a set of
dipoles (the projectile), the saturation effects encoded in the JIMWLK
equation are converted into unitarity effects in the evolution
of the scattering amplitudes. The latter are thus found to
obey an infinite hierarchy of evolution equations originally
derived by Balitsky \cite{bal}.

In what follows, we shall need only the first equation in this
hierarchy, which applies when the projectile is a single dipole.
This equation is most easily obtained by using the rapidity
increment $dY$ to accelerate the projectile (the dipole of size
$r$). Within the rapidity interval $dY$, either the dipole does not
split, in which case its scattering amplitude $T(r)\equiv T(r,r_0)$
remains unchanged, or it splits, in which case $T(r)$ is replaced by
the scattering amplitude of the two child dipoles. This leads to the
following evolution law
\begin{equation}
T(r)|_{Y+dY}=
\left\{
\begin{aligned}
T(r)|_Y\phantom{TTTTTTTT}&
\ \ \mbox{with probability}\ 1\!-\!\lambda\, \bar\alpha dY\\
T(z)+T(r\!-\!z)-T(z)T(r\!-\!z)|_Y&\ \ \mbox{with probability}\ \lambda\,
\bar\alpha dY
\end{aligned}
\right.
\label{law}
\end{equation}
where $z$ is distributed according to $p(z,r\!-\!z|r)\,d^2 z$.
Taking the limit $dY\rightarrow 0$
and replacing $\lambda$ and $p$ from Eqs.(\ref{lambda}),(\ref{p}),
one gets%
\footnote{The impact parameter dependence could
be easily put back in Eq.~(\ref{balitsky}).
We have omitted it for simplicity and since it is enough
for our purpose
to assume locality of the evolution.}
\begin{multline}
\partial_Y \langle T(r)\rangle_Y
=\frac{\bar\alpha}{2\pi}\int d^2 z \frac{r^2}{z^2(r\!-\!z)^2}
\big(
\langle T(z)\rangle_Y+\langle T(r\!-\!z)\rangle_Y
-\langle T(r)\rangle_Y\\
-\langle T(z)T(r\!-\!z) \rangle_Y \big).
\label{balitsky}
\end{multline}
As anticipated, Eq.~(\ref{balitsky}) is not a closed equation for $\langle T
\rangle$: it depends upon the correlator $\langle
T(z)T(r\!-\!z)\rangle_Y$.  A mean field approximation $\langle
T(z)T(r\!-\!z)\rangle\simeq \langle T(z)\rangle\langle
T(r\!-\!z)\rangle$ would cast Eq.~(\ref{balitsky}) into a closed form,
known as the Balitsky-Kovchegov (BK) equation \cite{bal,Kov}. The linearized
form of Eq.~(\ref{balitsky}) is recognized as the (dipole
version of) BFKL equation \cite{BFKL}.

\begin{figure}
\begin{center}
\epsfig{file=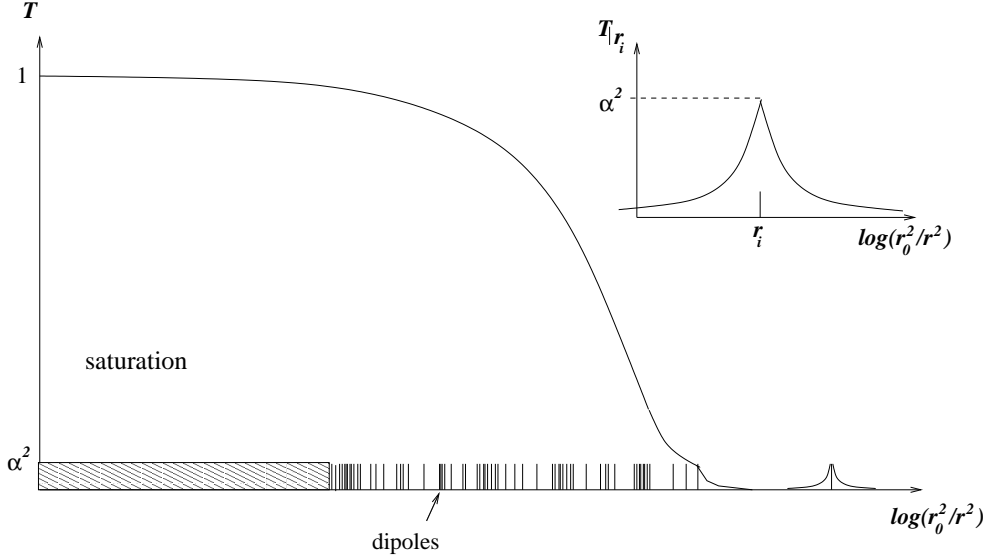,width=13cm} \caption{ The amplitude $T$ for a
typical partonic realization as a function of
$\rho=\ln(r_0^2/r^2)$. The individual dipoles seen at impact
parameter $b$ are represented by a short vertical line. The
straight line is the sum of their contributions to the amplitude.
In the saturation regime, the dipole description breaks down, that
is indicated by the filled box. {\it Upper right corner:} the
contribution of a single dipole to $T$,
Eq.~(\ref{Tel}).}\label{Fig1}
\end{center}
\end{figure}

Let us finally discuss the typical shape of $T(r)$ as resulting from
the previous considerations. It is a well known characteristic of the
BFKL evolution that the most probable splittings are those
in which both child dipoles have a size comparable to the size of
the parent dipole (this can be also checked on Eq.~(\ref{balitsky})).
Thus, if one starts with one dipole $r_0$ at $Y=0$, then the main
mechanism for the rise of $T(r)$ with $Y$ is a growing diffusion around the
size of the initial dipole $r_0$.  Consequently, in a typical partonic
configuration as obtained after a sufficiently large rapidity
evolution, the dipoles appear to be densely distributed around the
size $r_0$ (where $T(r)$ is large), but they become more rare with
decreasing $r$ (or increasing $\rho$), and for sufficiently large
$\rho$ one meets only rare fluctuations which involve one (or few)
dipoles and for which $T(r)\simeq\alpha_s^2$.  The typical partonic
realization is shown in Fig.~\ref{Fig1}: it is a front which with
increasing $Y$ progresses towards larger values of $\rho$. We define
the saturation scale $Q_s(Y)$ of a given partonic configuration by the
position of this front, that is, by the value of the inverse dipole
size for which $T$ reaches some predefined number $T_0$ of order one:
$T(1/Q_s(Y))=T_0$. We also define $\rho_s(Y)=\log(r_0^2 Q_s^2(Y))$.


\section{The energy dependence of the saturation scale}

When $T(r)$ is of order $\alpha_s^2$, the number of dipoles
participating in the scattering is small, and fluctuations dominate
the dynamics of $T(r)$.  By contrast, when $T(r)\gg\alpha_s^2$, the
fluctuations $\delta T$ in $T$ become relatively unimportant (since
typically $\delta T\sim \alpha_s\sqrt{T}$), so the dynamics is
self--averaging (for each individual front realization), and the
mean field description of a given event becomes justified.
Consequently, the evolution of $T$ in the bulk of the front
$T(r)\gg\alpha_s^2$ is essentially given by a mean field equation
(the BK equation). It turns out that, for the purpose of computing
the asymptotic energy dependence of the saturation momentum, one can
still rely on a {\it modified} mean field approximation, which is
obtained by introducing a factor $\Theta(T-\alpha_s^2)$ in the BK
equation. The latter implements the fact that, in a real event, in
which the occupation number is discrete, $T$ cannot become less than
$\alpha_s^2$ (cf. the discussion in Sec.~2).

To appreciate the dynamical role of this cutoff, it is useful to
notice an essential difference between the tail of a real event, and
that of the solution to the BK equation: whereas the front generated
by the BK equation has an exponential tail which extends up to
arbitrarily large $\rho$ (see Eq.~(\ref{front}) below), a real
event, on the other hand, is like a histogram whose front is
necessarily compact: for any $Y$, there exists a {\it foremost
occupied bin} (f.o.b.) $\rho_{\rm f.o.b.}\equiv \rho_{\rm
f.o.b.}(Y)$ such that $T(\rho_{\rm f.o.b.})|_Y > 0$ and $T(\rho)|_Y
= 0$ for any $\rho >\rho_{\rm f.o.b.}$. This implies that the
mechanism for front propagation is different in the two cases. For
the mean field approximation, the dominant mechanism is the {\it
local growth} within the tail of the distribution: at any $\rho \gg
\rho_s(Y)$, the local amplitude rises very fast due to the BFKL
instability, thus ``pulling" the front towards the right. By
contrast, in a real event, the local growth is not possible in the
empty bins on the right of the f.o.b., so the only way for the front
to progress into those bins is via {\it diffusion}, i.e. via
radiation from the occupied bins at $\rho <\rho_{\rm f.o.b.}$. 
But since diffusion is less effective than
the local growth, we expect the ``velocity'' $d\rho_s(Y)/dY$ of the
front (i.e., the exponential growth rate for the saturation
momentum) to be reduced in the real event as compared to the
corresponding prediction of the BK equation.

A simple way to try and capture this physical situation in
mathematical terms is to insert a cutoff $\Theta(T-\alpha_s^2)$ on
the growth term in the BK equation, while allowing the diffusion
there to remain operative even at arbitrarily small $T$. (This is
possible after separating the local growth term from the diffusion
term in the BFKL kernel with the help of a ``diffusion
approximation"; see, e.g., \cite{Mun} for details.) This simple
recipe was in fact invented by Brunet and Derrida \cite{Bru} in the
context of statistical physics.  They studied the propagation of
fronts in the presence of fluctuations associated with discreteness
in a variety of physical situations (see Ref. \cite{Pan} for a
review).  Using the mean field approximation to their dynamical
equations together with a suitable cutoff, they were able to compute
analytically the time evolution of the position of the front and of
its bulk shape. Although there is so far no full mathematical
justification of this procedure, it has been checked (through
numerical calculations) that it yields indeed the right value for
the velocity of the front at large times and for a large number of
particles.  This result is likely to be valid for all models that
fall in the universality class of the stochastic FKPP equation
\cite{Pan}.

We expect that, also in the present QCD context, that recipe give
the right asymptotics of $\rho_s(Y)=\ln (r_0^2 Q_s^2(Y))$ for
$\ln(1/\alpha_s^2)\gg 1$. Indeed, when viewed in the way exposed in
Sec.~2, the rapidity evolution of $T$  is essentially the same as
the time evolution of say the particle number density of a system
made of a number $N\sim 1/\alpha_s^2$ of diffusing and interacting
particles, and hence belongs to the class of models studied by
Brunet and Derrida.

We now turn to the practical computation of the rapidity dependence
of the saturation scale. Since it amounts to solving the BK equation
supplemented by a cutoff, it is useful to recall first the
asymptotic solutions to the BK equation without a cutoff and the way
how they set in (see e.g. \cite{Mun} for details). As the BK
equation falls into the universality class of the
Fisher-Kolmogorov-Petrovsky-Piscounov (FKPP) equation \cite{KPP},
its asymptotic solution is a traveling wave, i.e. a uniformly
translating front whose ``position'' is characterized by the
logarithm of the saturation scale $\rho_s(Y)$. It moves with the
``velocity'' $d\rho_s(Y)/dY=\bar\alpha\chi(\gamma_0)/\gamma_0$
towards smaller dipole sizes. Here,
$\chi(\gamma)=2\psi(1)-\psi(\gamma)-\psi(1\!-\!\gamma)$ is the
$\rho$--moment of the dipole splitting
probability~(\ref{splitting}), or, equivalently, the characteristic
function of the BFKL kernel, and $\gamma_0=0.6275...$ solves
$\chi'(\gamma_0)=\chi(\gamma_0)/\gamma_0$ \cite{Gri}. The front
exhibits a universal tail:
\begin{equation}
T(\rho,Y)\sim e^{-\gamma_0(\rho-\rho_s(Y))} \ \ \mbox{for}\ \
\rho-\rho_s\gg 1\ . \label{front}
\end{equation}
These asymptotics
set in diffusively
and spread over the range $\rho-\rho_s(Y)$ within the time interval
\begin{equation}
\bar\alpha \Delta Y\sim \frac{(\rho-\rho_s)^2}{2\chi''(\gamma_0)}\ .
\label{diffusiont}
\end{equation}
This process induces corrections to the asymptotic
$Y$-dependence of the saturation scale of the form
\begin{equation}
\frac{d\rho_s(Y)}{dY}=\bar\alpha \frac{\chi(\gamma_0)}{\gamma_0}
-\frac{3}{2\gamma_0}\frac{1}{Y}\ .
\label{satscal0}
\end{equation}
Note that the velocity and the shape of the front for
$\rho\gg\rho_s$ are completely determined by the linearized (BFKL)
equation, and do not depend on the exact form of the
nonlinearities: this is a very important consequence of the nature
of the propagation of the front, which is pulled along by its
tail. It implies that for a number of physical quantities, such as
$\rho_s$, we do not need to know the precise nonlinear mechanism
that enforces unitarity of $T$. This means in particular, that the
dipole picture is good enough for our purpose, although it is
incomplete.

Equation~(\ref{satscal0}) shows that the velocity of the front
increases with $Y$ up to its asymptotic value, which corresponds to
the velocity of a front which has the shape~(\ref{front}) all the
way down to $\rho\rightarrow\infty$. However, coming back to the
physical situation where, due to the discreteness of the dipole
number, the real front is a histogram, we see that $T$ can assume
the shape~(\ref{front}) only down to $T\sim \alpha_s^2$. Starting
from the initial condition at $Y=0$ and evolving it up to rapidity
$Y$, the amplitude first grows until it reaches the unitarity limit
$T=1$ around $r\sim r_0$. Then the traveling wave front forms, and
spreads from the point $\rho_s(Y)$ where $T\sim 1$ down to the point
$\rho$ at which $T\sim\alpha_s^2$. From Eq.~(\ref{diffusiont}) and
from the shape of the asymptotic front Eq.~(\ref{front}), the latter
process occurs within the rapidity interval
\begin{equation}
\bar\alpha\Delta Y= c\frac{\big[\ln(1/\alpha_s^2)/\gamma_0\big]^2}
{2\chi''(\gamma_0)} \label{time}
\end{equation}
 ($c$ is a number of order one)
during which the velocity of the front keeps increasing according
to Eq.~(\ref{satscal0}). But once the point where
$T\!\sim\!\alpha_s^2$ is reached, the front cannot extend to even
larger values of $\rho$ (corresponding to lower values of $T$), at
variance with the pure mean field case implemented by the BK equation.
Accordingly, the front velocity cannot increase anymore. Thus the
asymptotic $Y$-dependence of the saturation scale is
\begin{equation}
\frac{d\rho_s(Y)}{dY}=\bar\alpha\frac{\chi(\gamma_0)}{\gamma_0}
-3c\,\bar\alpha
\frac{\gamma_0\chi''(\gamma_0)}{\ln^2(1/\alpha_s^2)}\ .
\label{satscal}
\end{equation}
The calculation of $c$ requires a proper account of the
exact shape of the front, and yields $c=\pi^2/6$ \cite{Bru,Sho}.

The result~(\ref{satscal}) is identical to the one obtained in the
mean field approach of Ref.~\cite{Sho}, where the effects of
discreteness have been simulated by introducing a second barrier on
the phase--space for BFKL evolution (in addition to the first
barrier meant to enforce saturation \cite{Mue}), whose role looks
indeed very similar to that of our above cutoff on the value of $T$.
What was different, however, was the understanding of the physical
role played by this barrier: in Ref.~\cite{Sho}, the second barrier
was merely intended to impose unitarity on those evolution paths
which involve a single intermediate point. But it has not been
realized there that, for the evolutions leading to a final amplitude
$T_f\sim 1$ (as relevant for a study of saturation), the constraint
implied by the second barrier is actually sufficient to guarantee
unitarity for {\it arbitrary} paths (i.e. for paths with an
arbitrary number of intermediate points). That is, the generality of
Eq.~(\ref{satscal}) as being the correct result in the limit where
$Y\to \infty$ and $\alpha_s\to 0$ has not been recognized, nor
argued, in Ref.~\cite{Sho}.


\section{The physical amplitude and its scaling}

\begin{figure}
\begin{center}
\epsfig{file=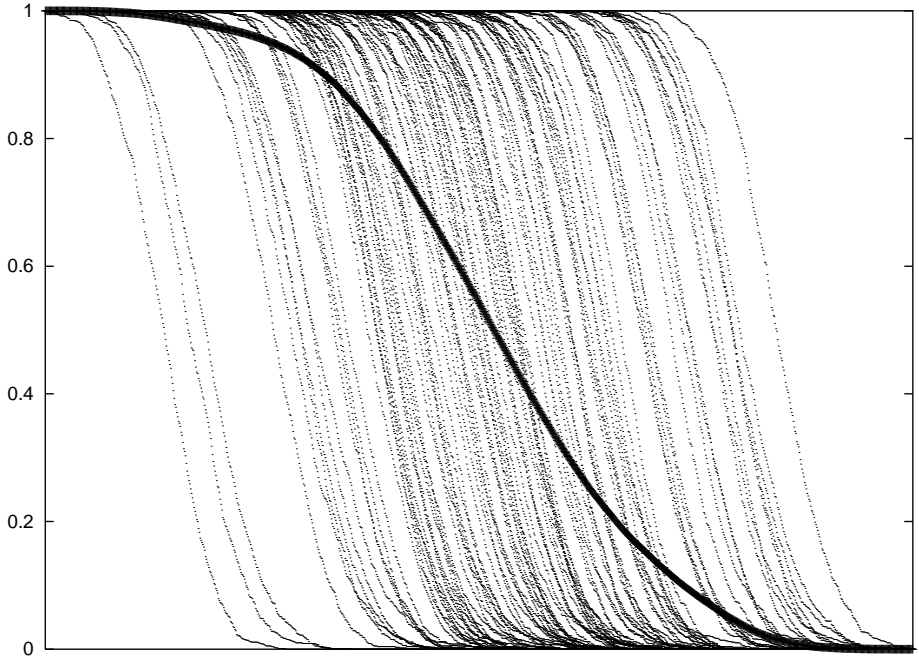,width=13cm} \caption{The scattering
amplitude $T$ for different partonic realizations at a given
rapidity against $\rho=\ln(r_0^2/r^2)$. The thick line is the
average over all realizations, i.e. the physical amplitude $A$,
see Eq.~(\ref{average}).} \label{Fig2}
\end{center}
\end{figure}

So far, we have followed the evolution of the amplitude $T(r)$
corresponding to one given partonic realization $|\omega\rangle$.
Each such realization undergoes a stochastic evolution given by
Eq.~(\ref{law}). At each rapidity, $T(r)$ has the universal
shape~(\ref{front}), up to fluctuations concentrated in its tail
$T\sim\alpha_s^2$. The position $\rho_s(Y)$ of the front exhibits
the $Y$--dependence\footnote{%
It is interesting to note in this
context that, for a given front realization, the asymptotic
velocity, Eq.~(\ref{satscal}), is reached {\it exponentially} fast
in $Y$ \cite{Pan}, at variance with the mean field problem, where
the corresponding approach is only power--like, cf.
Eq.~(\ref{satscal0}).} given by Eq.~(\ref{satscal}). However, since
the actual evolution is stochastic, the saturation scale $\rho_s$
undergoes a random walk about its mean $\langle\rho_s\rangle$, and
$\rho_s$ gets a variance $\sigma$. Physically, the origin of this
phenomenon can be traced to the statistical fluctuations in the tail
$T\sim\alpha_s^2$ of the amplitude: in the course of the evolution,
an unusually large number of dipoles may be created, which, after
further rapidity evolution, would ``pull'' the whole front ahead of
its typical evolution, resulting in a saturation scale $\rho_s$ for
this particular realization larger than the mean
$\langle\rho_s\rangle$. This mechanism leads to a spread of the
saturation scales of different partonic realizations, while the
shape of $T$ in its bulk remains identical. In particular, the
velocity of the average front $d\langle\rho_s \rangle/dY$ is still
given by Eq.~(\ref{satscal}) (at least, for sufficiently large $Y$;
see below), because this is the (asymptotic) velocity of all the
individual fronts making up the statistical ensemble. This is
illustrated in Fig.~\ref{Fig2}, which in fact applies to the
discrete statistical model in Ref.~\cite{Bru} (but a similar situation is
expected in QCD): the profile of the amplitude for partonic
realizations obtained from different stochastic evolutions over the
same rapidity interval are shown. The dispersion of $\rho_s$ is
manifest, as well as the universality of the shape of $T$.

The physical amplitude $A(r,Y)$ is obtained by averaging $T(r)$
over all Fock states at rapidity $Y$. In order to perform this average,
the expression for the variance $\sigma$ of the saturation scale
is needed.

The latter was recently studied in the statistical physics context for various
reaction-diffusion like models involving a finite number $N$ of
particles. The variance of the position of the front was seen to scale like\footnote{%
Formula~(\ref{sigma}) is borrowed from the third reference of \cite{Bru}, Eq.~(4), with
the replacement
$\sigma^2\leftrightarrow D_N\times t$. The time $t$ has to be replaced by $\bar\alpha Y$
and the number of particles $N$ is the maximum number of dipoles
of a given size $1/\alpha_s^2$.}
\begin{equation}
\sigma^2 = \langle\rho_s^2\rangle - \langle\rho_s\rangle^2
\sim {\bar{\alpha}Y\over \ln^3 (1/\alpha_s^2)}
\label{sigma}
\end{equation}
from numerical simulations~\cite{Bru}.
Although to our knowledge there is still no general analytical proof
of this result and the status of~(\ref{sigma}) is still that of a conjecture,
such behavior has been
checked in independent numerical work (see e.g. Ref.~\cite{Moro})
and is also likely to be very general~\cite{Pan}.
We also see on Eq.~(\ref{sigma}) that the fluctuations of $\rho_s$,
that are of order $\sigma\propto\sqrt{\bar\alpha Y}$, are indeed subleading
in $Y$ with respect to the effects of discreteness discussed in Sec.~3
(of order $\bar\alpha Y$).
This is a consistency check of the mean field calculation
used to obtain $\langle\rho_s\rangle$, see Eq.~(\ref{satscal}) .

Knowing the shape~(\ref{front}) of $T$, the value of the saturation
scale $\rho_s$ and the amplitude of its fluctuations $\sigma$, we
are in a position to evaluate the physical scattering amplitude
$A(\rho,Y)$. Up to higher moments of the distribution of $\rho_s$,
$A(\rho,Y)$ is obtained from the amplitudes $T(\rho)|_Y$ for each
particular realization of the Fock state of the target at rapidity
$Y$ (note that $T(\rho)|_Y$ is implicitly a function of $\rho_s$, as
manifest e.g. in Eq.~(\ref{front})), after averaging over the
corresponding saturation momenta with a Gaussian weight of variance
$\sigma$ (see also Fig.~\ref{Fig2}):
\begin{equation}
A(\rho,Y)=\frac{1}{\sigma\sqrt{2\pi}} \int d\rho_s\, T(\rho)|_Y\,
\exp\left( -\frac{(\rho_s-\langle\rho_s\rangle)^2} {2\sigma^2}
\right)\ . \label{average}
\end{equation}
We deduce the following scaling form for the physical amplitude:
\begin{equation}
A(\rho,Y)=A\left(\frac{\rho-\langle\rho_s(Y)\rangle}{\sqrt{\bar\alpha
Y/\ln^3(1/\alpha_s^2)}}\right)\ , \label{scaling}
\end{equation}
up to ${\mathcal O}(1/\sqrt{Y})$ corrections. It is obvious from
that formula that, at sufficiently high energies, geometric scaling
does not hold for the physical amplitude. This feature is a direct
consequence of the statistical nature of the parton model: the
statistical fluctuations of the number of dipoles translate into a
random wandering of the saturation scale, and after averaging over
partonic realizations, the scaling form~(\ref{scaling}) results.

The violation of geometric scaling manifest in Eq.~(\ref{scaling})
was already noted in
Ref. \cite{Sho}, however, the square root in the denominator of
the scaling variable was missing because of a lack of fluctuations
in the tail of the distribution: the approach used there was
relying on mean field throughout, missing the stochastic nature of the
evolution.

\section{Conclusion}

We have shown that high energy QCD is similar to a reaction-diffusion
problem, well studied by statistical physicists.
We have been able to obtain the $Y$-dependence of the saturation scale,
see Eq.~(\ref{satscal}),
confirming results obtained recently by different methods \cite{Sho}.
We have also derived the scaling form of
the asymptotic dipole-dipole scattering amplitude (\ref{scaling}), which
is related to the dispersion of saturation scales between
different ``events''
(corresponding to different partonic realizations).
That scaling is clearly not geometric.

A recurrent theme in this Letter has been universality
that guarantees that the lowest order results do not depend on
the details of the model. The exact way how saturation comes
about was not an issue, as well  the details of the elementary
dipole interaction do not enter the leading order results (in $Y\gg 1$ and
$\alpha_s^2\ll 1$) that we have obtained here.
Further terms in these expansions will be model dependent, and thus
much more difficult to get.

One of the points that remain to be studied is how fast the
computed asymptotics set in. The BK equation may still be a good
approximation for a large target (like a nucleus) and in the first
stages of the evolution, when the traveling wave front has not
diffused down to $T\sim\alpha_s^2$. More precise numerical studies
\cite{Sal,Rum} may help to clarify this point.

Finally, as mentioned in the Introduction, the dependence upon the
impact parameter plays an important role for the overall physical
picture of unitarization. This will be discussed at length
somewhere else \cite{lle}.


\section*{Acknowledgments}
S.M. warmly thanks Dr. E. Brunet and Pr. B. Derrida for
illuminating discussions about their work and its possible
application to QCD. He thanks 
Dr. G.~Salam for illuminating discussions, and in particular, for
pointing out the role of the impact parameter dependence.
He also acknowledges support from the
RIKEN-BNL Summer Program and from Columbia University at the time
when this project was being finalized. 
We thank Drs. R.~Peschanski, G.~Salam
and D.~N. Triantafyllopoulos for helpful discussions.


\end{document}